\documentclass{PoS}
\usepackage{amssymb,amsmath,amsthm,bm}

\title{Neutrino electromagnetic properties and new physics}

\ShortTitle{Neutrino and new physics}

\author{\speaker{Alexander Studenikin}\\
        Author affiliation\\
        Department of Theoretical Physics, Faculty of Physics, Moscow State University, Moscow 119991, Russia\\
        Joint Institute for Nuclear Research, Dubna 141980, Moscow Region, Russia\\
        E-mail: \email{studenik@srd.sinp.msu.ru}}

\author{Ilya Tokarev\\
        Affiliation\\
        Department of Theoretical Physics, Faculty of Physics, Moscow State University, Moscow
        119991, Russia\\
        E-mail: \email{tokarev.ilya.msu@gmail.com}}

\abstract{
New effects of nontrivial neutrino electromagnetic properties are investigated on the basis of exact solutions of the modified Dirac equations for neutrinos in dense magnetized rotating matter. The effect of spatial separation of different types of neutrinos and antineutrinos (different in flavors and energies) moving inside dense magnetized rotating matter is predicted. We also describe a new mechanism of a star rotation frequency shift by neutrinos escaping the star (termed ``Neutrino Star Turning'' mechanism, $\nu S T$). The $\nu S T$ mechanism can explain the origin of pulsar ``anti-glitches'' and ordinary glitches as well. Considering a possible effect of the $\nu S T$ mechanism on speeding up of a pulsar together with the observational data on pulsars, we obtain a new limit on a neutrino millicharge $q_{\nu}<1.3\times10^{-19}e_0$ that is one of the strongest limits on this value obtained in astrophysics.
}

\FullConference{The European Physical Society Conference on High Energy Physics -EPS-HEP2013\\
		18-24 July 2013\\
		Stockholm, Sweden}

\begin{document}

\section{Introduction}

Neutrino electromagnetic properties are among the most intriguing and exciting problems in modern particle physics. Within the Standard Model in the limit of massless neutrinos the particle electromagnetic properties vanish. However, in different extensions of the Standard Model a massive neutrino has nontrivial electromagnetic properties (for a review of the neutrino electromagnetic properties see \cite{Giunti:2008ve,Broggini:2012df}). That is why it is often claimed that neutrino electromagnetic properties open ``a window to the new physics" \cite{Studenikin:2008bd}.

A neutrino magnetic moment, as expected in the easiest generalization of the Standard Model, is very small and proportional to the neutrino mass $m$: $\mu_\nu \approx 3 \times 10^{-19}\mu_B(m/1~{\rm eV})$ \cite{Fujikawa:1980yx}, where $\mu_B$ is the Bohr magneton. Much greater values are predicted in other various Standard Model generalizations (for details see \cite{Giunti:2008ve,Broggini:2012df}).There is a set of theoretical models with the absence of hypercharge quantization \cite{Foot} where neutrinos may have also a nonzero electric millicharge. The most severe experimental constraints on the magnetic moment and neutrino millicharge are $\mu_{\nu}\leq2.9\times10^{-11}\mu_B$ \cite{Beda:2012zz} and $q_0\leq10^{-21}e_0$ \cite{Marinelli:1983nd,Baumann:1988ue} correspondingly. In astrophysics the best model-independent limits are $\mu_{\nu}\leq 3.2\times10^{-11}\mu_B$ \cite{Raffelt} and $q_0\leq3\times10^{-17}e_0$ \cite{Raffelt,Barbiellini:1987zz}.

In this short notes we have further developed an advanced approach \cite{StudTern,Stud2008} of the nontrivial neutrino electromagnetic properties investigation that is based on the method of exact solutions of modified Dirac equations for description neutrino quantum states in a background environment (dense matter and external electromagnetic fields). Within this method we have obtained a new solution of the Dirac equation describing a millicharged neutrino with anomalous magnetic moment in dense magnetized matter as well as the solution accounting for the effect of matter rotation. On this basis we have predicted a new effect of a deflection of millicharged neutrinos propagating inside a dense magnetized rotating matter. This new effect can explain the absence of joint observations of initially coincided light and neutrino signals from astrophysical transient sources in recent terrestrial experiments. We have also predicted a new mechanism of a star rotation frequency shift originated by neutrinos escaping the star, the ``Neutrino Star Turning'' ($\nu ST$) mechanism. The $\nu ST$ mechanism can explain the origin of pulsar glitches and ``anti-glitches'' as well. From the demand that one has to avoid contradictions of possible effect of a pulsar seed up due to the $\nu ST$ mechanism with the observational data on pulsars, we have obtained a new limit on a neutrino millicharge $q_{\nu}<1.3\times10^{-19}e_0$. This is, in fact, one of the strongest astrophysical limits on the neutrino millicharge.

\section{Millicharged neutrino with anomalous magnetic moment in dense magnetized matter}

A millicharged neutrino with a nonzero magnetic moment moving in an external magnetic field and dense matter composed of neutrons, electrons and protons is described by the modified Dirac equation
\begin{equation}
\label{dirac}
\left(\gamma_{\mu}P^{\mu}-\frac12\gamma_{\mu}(1+\gamma_5)f^{\mu}
-\frac{i}{2}\mu\sigma_{\mu\nu}F^{\mu\nu}-m\right)\Psi(x)=0,
\end{equation}
where $P^{\mu}=p^{\mu}+q_0A^{\mu}$ is a particle kinetic momentum, $q_0$ is an absolute value of the neutrino millicharge ($q_{\nu}=-q_0$), $F^{\mu\nu}=\partial^{\mu}A^{\nu}-\partial^{\nu}A^{\mu}$ with $A^{\mu}=(0,-\frac{yB}{2},\frac{xB}{2},0)$ describes the constant magnetic field coincided with the third coordinate axis $\bm e_z$ and $\mu$ is the neutrino anomalous magnetic moment. Matter potential $V_m=\frac12\gamma_{\mu}(1+\gamma_5)f^{\mu}$ is originated by neutrino weak interactions with the background matter and the explicit form of $f^{\mu}$ depends on the background particles densities, velocities and polarizations \cite{StudTern,Balantsev:2010zw}. In case of unpolarized static matter $f^{\mu}=Gn(1, 0, 0, 0)$, where $G=\frac{G_F}{\sqrt{2}}$ ($G_F$ is the Fermi constant) and $n=-n_n+n_p(1-4\sin^2\theta_w)+n_e(\rho+4\sin^2\theta_w)$ is effective matter density ($\rho=1$ for an electron neutrino and $\rho=-1$ for muon and tau neutrinos).

Eq.~(\ref{dirac}) can be written in the Hamiltonian form $i\frac{\partial}{\partial t}\Psi(x)=H\Psi(x)$ with the Hamiltonian
\begin{equation}
\label{hamiltonian}
H=\gamma_0\bm{\gamma P}+\gamma_0m+\gamma_0\sigma_3\mu
B+(1+\gamma_5)\frac{Gn}{2}.
\end{equation}
The exact solution for the neutrino wave function can be obtained in the form
\begin{equation}
\label{solution}
\Psi(x)=
\sqrt{\frac{q_0B}{2\pi L}}e^{-i(p_ot-p_3z)}
\begin{pmatrix}
C_1\mathcal{L}_s^{l-1}(\frac{q_0B}{2}r^2)e^{i(l-1)\varphi} \\
iC_2\mathcal{L}_s^l(\frac{q_0B}{2}r^2)e^{il\varphi} \\
C_3\mathcal{L}_s^{l-1}(\frac{q_0B}{2}r^2)e^{i(l-1)\varphi} \\
iC_4\mathcal{L}_s^l(\frac{q_0B}{2}r^2)e^{il\varphi}
\end{pmatrix},
\end{equation}
where $\mathcal{L}_s^l(\frac{q_0B}{2}r^2)$ are the Laguerre functions ($N=l+s=0,1,2...$).

To advance in the exact solution of Eq.~(\ref{dirac}) one should determine the spin coefficients $C_i$. Note that spin properties of a particle in simultaneous presence of matter and magnetic field have never been described before. To this purpose we introduce a new type of the spin operator
\begin{equation}
\label{spin_op}
S=S_{tr}\cos\alpha-S_{long}\sin\alpha, \quad
\sin\alpha=\frac{Gn}{\sqrt{(Gn)^2+(2\mu B)^2}},
\end{equation}
which is the weighted superposition of the operators of longitudinal and transverse polarizations
\begin{equation}
S_{long}=\frac{\bm{\Sigma P}}{m}, \quad
S_{tr}=\Sigma_3+\frac{i}{m}
\begin{pmatrix}
0 & -\sigma_0 \\
\sigma_0 & 0 \\
\end{pmatrix}
[\bm\sigma\times\bm P]_3,
\end{equation}
where $\sigma_{\mu}$ are the Pauli matrixes. The spin operator~(\ref{spin_op}) commutes with the Hamiltonian~(\ref{hamiltonian}) and yields the spin integral of motion
\begin{equation}
\label{spin_spectrum}
S=\frac{\zeta}{m}\sqrt{(m\cos\alpha-p_3\sin\alpha)^2+2Nq_0B}, \quad \zeta=\pm 1.
\end{equation}
Using the spin integral of motion~(\ref{spin_spectrum}) one can obtain the energy spectrum of the Hamiltonian~(\ref{hamiltonian})
\begin{equation}
\label{energy_spectrum}
p_0=\frac{Gn}{2}+\varepsilon\sqrt{
p_3^2+2Nq_0B+m^2+\left(\frac{Gn}{2}\right)^2+(\mu B)^2+
2mS\sqrt{\left(\frac{Gn}{2}\right)^2+(\mu B)^2}}, \quad
\varepsilon=\pm1,
\end{equation}
and the spin coefficients
\begin{equation}
\label{coefficients}
\begin{aligned}
C_1&=\frac12
\sqrt{1+\frac{m\cos\alpha-p_3\sin\alpha}{mS}}
\sqrt{1+\sin(\alpha+\beta)},\\
C_2&=\frac12\delta_1\zeta
\sqrt{1-\frac{m\cos\alpha-p_3\sin\alpha}{mS}}
\sqrt{1+\sin(\alpha-\beta)},\\
C_3&=\frac12\delta_2
\sqrt{1+\frac{m\cos\alpha-p_3\sin\alpha}{mS}}
\sqrt{1-\sin(\alpha+\beta)},\\
C_4&=\frac12\delta_3\zeta
\sqrt{1-\frac{m\cos\alpha-p_3\sin\alpha}{mS}}
\sqrt{1-\sin(\alpha-\beta)}.\\
\end{aligned}
\end{equation}
We use the notations
$\delta_1=-\mathrm{sgn}[\sin\alpha+\cos\beta],\,
\delta_2=\mathrm{sgn}[\cos(\alpha+\beta)],\,\delta_3=\mathrm{sgn}[\cos\alpha+\sin\beta]$
and introduce a new angle $\beta$,
\begin{equation}
\label{beta}
\cos\beta=\frac{p_3\cos\alpha+m\sin\alpha}{p_0-\frac{Gn}{2}}.
\end{equation}

Note that Eqs.~(\ref{solution}) and (\ref{spin_spectrum})-(\ref{beta}) represent the exact solution of the modified Dirac equation (\ref{dirac}) \cite{Balantsev,Studenikin:2012vi} that describes the millicharged neutrino with the nonzero magnetic moment in the dense magnetized matter.

\section{Millicharged neutrino in dense magnetized rotating matter}

A millicharged neutrino in the external magnetic field and dense rotating matter is described by the modified Dirac equation
\begin{equation}
\label{quantum_equation}
\left(\gamma_{\mu}P^{\mu}-\frac12\gamma_{\mu}(1+\gamma_5)f^{\mu}-m\right)\Psi(x)=0,
\end{equation}
that can be easily obtained from Eq.~(\ref{dirac}) by reducing the term proportional to the magnetic moment $\mu$. Note that the modified Dirac equation in form~(\ref{quantum_equation}) for the first time was represented and solved in \cite{StudTern} for a neutrino with zeroth neutrino millicharge ($q_{\nu}=0$) in a static matter. The effect of transversally moving matter was accounted for in \cite{StudSavochkin}.

We have considered a new case of background environment that interconnects an external electromagnetic field and dense rotating matter when both matter rotation vector $\bm\omega$ and the magnetic field $\bm B$ are coincided with the third coordinate axis $\bm e_z$. In this case $f^{\mu}=Gn(1, -y\omega, x\omega, 0)$, where $\omega$ is a matter rotation frequency. The solution of Eq.~(\ref{quantum_equation}) represents discrete energy spectrum of a millicharged neutrino in the dense magnetized rotating matter \cite{Balantsev,Studenikin:2012vi}
\begin{equation}
\label{energy_spectrum_rot}
p_0=\sqrt{p_3^2+2N|q_0B-2Gn\omega|+m^2}+Gn,
\end{equation}
where $N=0,1,2..$ is a number of the modified Landau level. Within a quasi-classical interpretation
the neutrino discrete energy states~(\ref{energy_spectrum}) can be explained as a result of action of an effective force \cite{Stud2008} that is produced by both weak and electromagnetic interactions of a millicharged neutrino with the dense magnetized rotating matter and has the form
\begin{equation}
\label{force}
\bm F=-(q_0B-2Gn\omega)\left[\bm\beta\times\bm{e}_z\right],
\end{equation}
where $\bm\beta$ is a neutrino velocity. The effective force~(\ref{force}) exemplifies the interconnection of these two types of fundamental interactions and is not vanished even in the case of a zero neutrino electric charge. This force seems to be very weak for any reasonable choice of background parameters, but, nevertheless, can produce new effects that can be observed in terrestrial experiments \cite{Studenikin:2012vi}. In particular, during a supernova core collapse due to the action of the force~(\ref{force}) escaping neutrinos can be deflected on an angle
\begin{equation}
\Delta\phi \simeq \frac{R_S}{R}\sin\theta, \quad R=\sqrt{\frac{2N}{|q_0B-2Gn\omega|}},
\end{equation}
where $R_S$ is the radius of the star, $R$ is the radius of the neutrino trajectory and $\theta$ is an azimuthal angle of neutrino propagation. We predict that initially coincided light and neutrino beams will be spatial separated after passing through a dense rotating magnetized matter. Therefore in terrestrial experiments joint observations of initially coincided light and neutrino signals from an astrophysical transient source should not occur due to their spatial separation $\Delta L\simeq\Delta\phi L$ ($L$ is distance to the source). This new effect can explain the recent experimental results of the ANTARES experiment \cite{antares}.

On the other hand the feedback of the effective force~(\ref{force}) from the escaping neutrinos to the star should effect the star evolution. In particular, a torque produced by the escaping neutrinos shifts the star angular frequency
\begin{equation}
\label{delta_omega}
|\triangle\omega_0|=\frac{5N_{\nu}}{6M_{S}}|q_0B-2Gn\omega_0|,
\end{equation}
where $\triangle\omega_0=\omega-\omega_0$ ($\omega_0$ is an initial star rotation frequency), $M_S$ is the star mass and $N_{\nu}$ is the number of the escaping neutrinos. We have termed the phenomenon as the ``Neutrino Star Turning'' ($\nu ST$) mechanism. Note that depending on the neutrino millicharge sign the star rotation due to the $\nu ST$ mechanism can either spin up ($\triangle\omega_0>0$ for $q_{\nu}<0$) or spin down ($\triangle\omega_0<0$ for $q_{\nu}>0$).

In case of a zero neutrino millicharge the $\nu ST$ mechanism is produced only due to weak interactions and yields
\begin{equation}
\label{delta_omega_weak}
\frac{|\triangle\omega_0|}{\omega_0}\simeq10^{-8}.
\end{equation}
where we have considered the star with mass $M_S = 1.4  M_{\odot}$ ($M_{\odot}$ is the Solar mass) and $N_{\nu}=10^{58}$ escaped neutrinos with energy $\sim10$ MeV \cite{SN1987A}. The value of the relative rotation frequency shift~(\ref{delta_omega_weak}) is very close to sporadical sudden increase of a pulsar rotation frequency (a pulsar glitch \cite{glitch}). The obtained results are very important for astrophysics in light of the recent observed ``anti-glitch'' event \cite{antiglitch} that is sudden decrease of a pulsar rotation frequency. The $\nu ST$ mechanism can be used to explain both glitches and ``anti-glitches'' as well.

In case of a nonzero neutrino millicharge electromagnetic interactions provide the dominant contribution to the $\nu ST$ mechanism and one can neglect the weak interaction of neutrinos with background matter. From Eq.~(\ref{delta_omega}) we obtain
\begin{equation}
\label{delta_omega_nonzero_charge}
\frac{|\triangle\omega_0|}{\omega_0}=\frac{7.6q_0}{e_0}\times
10^{18}\left(\frac{P_0}{10\text{ s}}\right)
\left(\frac{N_{\nu}}{10^{58}}\right)
\left(\frac{1.4  M_{\odot}}{M_{S}}\right)
\left(\frac{B}{10^{14}\textrm{G}}\right),
\end{equation}
where $P_0$ is a pulsar initial spin period. The current pulsar timing observations \cite{pulsar_timing} show that the present-day rotation periods are up to $10$ s. However, the rotation during the life of a pulsar spins down due to several various mechanisms and dominantly due to a magnetic dipole braking. All of the estimations of feasible initial pulsars spin periods give the values that are very close to the present observed periods. Therefore, possible existence of a nonzero negative neutrino millicharge should not significantly speed up the rotation of a born pulsar (or speed down in case of a positive neutrino millicharge). From the straightforward demand $|\triangle\omega_0| < \omega_0$ and Eq.~(\ref{delta_omega_nonzero_charge}) we obtain the upper limit on the neutrino millicharge
\begin{equation}
\label{bound_q_nu}
q_0<1.3\times10^{-19} e_0.
\end{equation}
That is, in fact, one of the most severe astrophysical limits on the neutrino millicharge \cite{Raffelt}.

\acknowledgments
The work on this paper has been partially supported by the Russian Foundation for Basic Research (grant No. 14-02-31816 mol\_a) and by the Russian Science Foundation (grant No. 14-12-00033).

\end{document}